\documentclass[twocolumn,prb,showpacs,superscriptaddress,
  preprintnumbers, amsmath,amssymb]{revtex4}

\usepackage[dvips]{graphicx}
\usepackage{dcolumn}
\usepackage{bm}

\begin{document}


\title{Electronic structure of SrPt$_4$Ge$_{12}$: a combined
photoelectron spectroscopy and band structure study}

\author{H.~Rosner}
\affiliation{ Max-Planck-Institut f\"ur Chemische Physik fester
Stoffe, N\"othnitzer Stra{\ss}e 40, 01187 Dresden, Germany}

\author{J.~Gegner}
\affiliation{II. Physikalisches Institut, Universit\"{a}t zu
K\"{o}ln, Z\"{u}lpicher Stra{\ss}e 77, 50937 K\"{o}ln, Germany}

\author{D.~Regesch}
\affiliation{II. Physikalisches Institut, Universit\"{a}t zu
K\"{o}ln, Z\"{u}lpicher Stra{\ss}e 77, 50937 K\"{o}ln, Germany}

\author{W.~Schnelle}
\affiliation{ Max-Planck-Institut f\"ur Chemische Physik fester
Stoffe, N\"othnitzer Stra{\ss}e 40, 01187 Dresden, Germany}

\author{R.~Gumeniuk}
\affiliation{ Max-Planck-Institut f\"ur Chemische Physik fester
Stoffe, N\"othnitzer Stra{\ss}e 40, 01187 Dresden, Germany}

\author{A.~Leithe-Jasper}
\affiliation{ Max-Planck-Institut f\"ur Chemische Physik fester
Stoffe, N\"othnitzer Stra{\ss}e 40, 01187 Dresden, Germany}

\author{H.~Fujiwara}
\affiliation{II. Physikalisches Institut, Universit\"{a}t zu
K\"{o}ln, Z\"{u}lpicher Stra{\ss}e 77, 50937 K\"{o}ln, Germany}

\author{T.~Haupricht}
\affiliation{II. Physikalisches Institut, Universit\"{a}t zu
K\"{o}ln, Z\"{u}lpicher Stra{\ss}e 77, 50937 K\"{o}ln, Germany}

\author{T.~C.~Koethe}
\affiliation{II. Physikalisches Institut, Universit\"{a}t zu
K\"{o}ln, Z\"{u}lpicher Stra{\ss}e 77, 50937 K\"{o}ln, Germany}

\author{H.-H.~Hsieh}
\affiliation{Chung Cheng Institute of Technology, National Defense
  University, Taoyuan 335, Taiwan}

\author{H.-J.~Lin}
\affiliation{National Synchrotron Radiation Research Center (NSRRC),
  101 Hsin-Ann Road, Hsinchu 30077, Taiwan}

\author{C.~T.~Chen}
\affiliation{National Synchrotron Radiation Research Center (NSRRC),
  101 Hsin-Ann Road, Hsinchu 30077, Taiwan}

\author{A.~Ormeci}
\affiliation{ Max-Planck-Institut f\"ur Chemische Physik fester
  Stoffe, N\"othnitzer Stra{\ss}e 40, 01187 Dresden, Germany}

\author{Yu.~Grin}
\affiliation{ Max-Planck-Institut f\"ur Chemische Physik fester
  Stoffe, N\"othnitzer Stra{\ss}e 40, 01187 Dresden, Germany}

\author{L.~H.~Tjeng}
\affiliation{II. Physikalisches Institut, Universit\"{a}t zu K\"{o}ln,
   Z\"{u}lpicher Stra{\ss}e 77, 50937 K\"{o}ln, Germany}

\date{\today}

\hyphenation{ pho-to-emis-sion Lan-than-oid-spe-zi-fi-sche NSRRC}

\begin{abstract}
We present a combined study of the electronic structure of the superconducting
skutterudite derivative SrPt$_4$Ge$_{12}$ by means of X-ray photoelectron spectroscopy
and full potential band structure calculations including an analysis of the chemical
bonding. We establish that the states at the Fermi level originate predominantly from the
Ge 4$p$ electrons and that the Pt $5d$ shell is effectively full. We find excellent
agreement between the measured and the calculated valence band spectra, thereby
validating that band structure calculations in combination with photoelectron
spectroscopy can provide a solid basis for the modeling of superconductivity in the
compounds $M$Pt$_4$Ge$_{12}$ ($M$ = Sr, Ba, La, Pr) series.
\end{abstract}


\maketitle

\section{introduction}

Compounds with crystal structures featuring a rigid covalently
bonded framework enclosing differently bonded guest
atoms attracted much attention in the last decade. In particular
the skutterudite and clathrate families have been investigated in
depth, and a fascinating diversity of physical phenomena is
observed, many of which are due to subtle host-guest interactions.
Among the skutterudites they range from magnetic ordering to
heavy-fermion and non-Fermi liquids, superconductivity, itinerant
ferromagnetism, half-metallicity, and good thermoelectric
properties.
\cite{Uher01,SalesREHandbook,LeitheJasper03etal,Nolas99}
Superconductivity of conventional
\cite{Meisner81,Kawaji95,Tanigaki03} and heavy-fermion type is
found in skutterudites with $T_\mathrm{c}$'s up to 17\,K.
\cite{EDBauer02,Imai07a,Shirotani05}

The formula of the filled skutterudites, being derived from the
mineral CoAs$_3$, is given by $M_yT_4X_{12}$, with $M$ a cation, $T$ a
transition metal, and $X$ usually a pnicogen (P, As, or Sb). The $M$ atoms
reside in large icosahedral cages formed by [$TX_6$] octahedra. A new
family of superconducting skutterudites $M$Pt$_4$Ge$_{12}$ ($M$ = Sr,
Ba, La, Pr, Th) has been reported recently.
\cite{Bauer07a,Gumeniuk08a,Kaczorowski08} With trivalent La and Pr,
$T_c$'s up to 8.3\,K are observed. The compounds with the divalent
cations Sr and Ba have lower $T_c$'s around
5.0\,K.\cite{Bauer07a,Gumeniuk08a}

Band structure calculations predict that the electronic density of
states (DOS) close to the Fermi level $E_F$ is determined by Ge-$4p$
states in all $M$Pt$_4$Ge$_{12}$ materials.\cite{Bauer07a,Gumeniuk08a,Tran09}
The position of $E_F$ is adjusted by the electron count on the
polyanionic host structure.  This leads to the situation that the band
structure at $E_F$ can be shifted in an almost rigid-band manner by
``doping'' of the polyanion, which can be achieved either by charge
transfer from the guest $M$ or by internal substitution of the
transition metal $T$. Recently, this principle was demonstrated on the
Pt-by-Au substitution in BaPt$_{4-x}$Au$_x$Ge$_{12}$: while
BaPt$_4$Ge$_{12}$ has a calculated DOS of 8.8\,states/(eV f.u.),
at the composition BaPt$_3$AuGe$_{12}$ the DOS is enhanced to 11.5\,states/(eV f.u.).\cite{Gumeniuk08b} Experimentally, an increase
of the superconducting $T_c$ from 5.0\,K to 7.0\,K.\cite{Gumeniuk08b}
was observed. The rigid-band shift of the DOS peak at $E_F$ with gold
substitution is due to the Pt(Au) $5d$ electrons which, according to
band structure calculations, lie rather deep below $E_F$ and provide
only a minor contribution to the DOS at $E_F$.\cite{Gumeniuk08b}

Another prediction from the band structure calculations concerns the
special role played by the Pt $5d$ states in SrPt$_4$Ge$_{12}$ for the
chemical bonding. It is known, assuming two-center-two-electron bonds
within the $T$-$X$ framework for the binary skutterudites, that 72
electrons are required for the stabilization of the [$T$$_4$$X$$_{12}$]
formula unit.\cite{Uher01}
In the case of SrPt$_4$Ge$_{12}$, the total number of $s$ and $p$
electrons of Sr, Pt and Ge is 2+4$\times$2+12$\times$4\,=\,58. To
achieve the target value of 72, each Pt atom should use 3.5 $d$ electrons
for bonding, which would be a rather large value compared to the one
$d$ electron per Co atom in Co$_4$As$_{12}$ (4$\times$CoAs$_3$).

\begin{figure}[t,b]
     \includegraphics[width=0.40\textwidth]{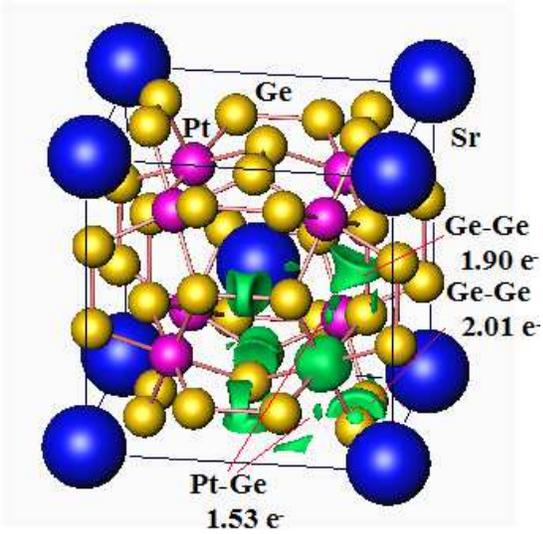}
          \vspace{-2mm}
     \caption{Chemical bonding in SrPt$_4$Ge$_{12}$: isosurface of ELI
       revealing Ge-Ge and Pt-Ge bonds together with their electron
       counts.  }
    \label{theo_elf}\vspace{-3mm}
\end{figure}

In order to determine the electron counts for SrPt$_4$Ge$_{12}$ as a
characteristic for the chemical bonding, we evaluated the so-called
electron localizability indicator (ELI).  The combined analysis of the
ELI and electron density (ED), see Figure~\ref{theo_elf}, shows indeed
three types of attractors in the valence region: two representing
Ge-Ge bonds and one reflecting the Pt-Ge bond. No attractors were
found between Sr and the framework atoms, suggesting predominantly
ionic interaction here. The Ge-Ge bonds are two-electron bonds
(electron counts 1.90 and 2.01), the Pt-Ge bond has a count of only
1.53 electrons, summing up to 60.18 electrons total per
[Pt$_4$Ge$_{12}$] formula unit.
This means that only about 0.5 $d$ electrons per Pt are participating
in the stablization of the [Pt$_4$Ge$_{12}$] polyanion. Both
procedures, valence electron counting and combined ELI/ED analysis
yield unusual results and raise the question about the role of 5$d$
electrons of Pt in the formation of the $M$Pt$_4$Ge$_{12}$ compounds.

Up to now, no spectroscopic data are available to challenge the above
mentioned band structure predictions and chemical bonding
analysis. Such a validation is important since one would like to know
whether band theory can provide a solid basis for the modeling of
superconductivity in the $M$Pt$_4$Ge$_{12}$ ($M$ = Sr, Ba, La, Pr)
series. We therefore set out to perform a comparative study of the
valence band electronic structure of the superconducting skutterudite
derivative SrPt$_4$Ge$_{12}$ by means of x-ray photoelectron
spectroscopy (PES) and full potential band structure calculations.

\section{methods}

Samples were prepared by standard techniques as described in Refs.
\onlinecite{Gumeniuk08a} and \onlinecite{Gumeniuk08b}.  Metallographic
and electron microprobe tests of polished specimens detected only
traces of PtGe$_2$ ($<$\,4\,vol{\%}) and SrPt$_2$Ge$_2$
($<$\,1\,vol{\%}) as impurity phases in the sample SrPt$_4$Ge$_{12}$.
EPMA confirmed the ideal composition of the target phase. The lattice
parameter is 8.6509(5)\,\AA , as reported earlier.  \cite{Gumeniuk08a}
For the cation position full occupancy was derived from full-profile
crystal structure refinements of powder XRD data, which are in good
agreement with single crystal data obtained in
Ref.~\onlinecite{Bauer07a}.

The PES experiments were performed at the Dragon beamline of the NSRRC
in Taiwan using an ultra-high vacuum system (pressure in the low
10$^{-10}$ mbar range) which is equipped with a Scienta SES-100
electron energy analyzer. The photon energy was set to 700 eV and to
190 eV. The latter energy is close to the Cooper minimum in the
photo-ionization cross section of the Pt $5d$ valence
shell.\cite{Yeh85} The overall energy resolution was set to 0.35\,eV
and 0.25\,eV, respectively, as determined from the Fermi cut-off in
the valence band of a Au reference which was also taken as the zero of
the binding energy scale. The 4$f_{7/2}$ core level of Pt metal was
used as an energy reference, too. The reference samples were scraped
\textit{in-situ} with a diamond file. The polycrystalline
SrPt$_4$Ge$_{12}$ sample with dimensions of $2$$\times 2$$\times
3$\,mm$^3$ was cleaved \textit{in-situ} exposing a shiny surface and
measured at room temperature at normal emission.

Electronic structure calculations within the local density
approximation (LDA) of DFT were employed using the full-potential
local-orbital code FPLO (version 5.00-19) .\cite{FPLOKoepernik99}  In
the full-relativistic calculations, the exchange and correlation
potential of Perdew and Wang \cite{PerdewWang92} was used. As the
basis set, Sr(4$s$, 4$p$, 5$s$, 5$p$, 4$d$), Pt (5$s$, 5$p$, 6$s$,
6$p$, 5$d$), and Ge (3$d$, 4$s$, 4$p$,4$d$) states were
employed. Lower-lying states were treated as core. A very dense
$k$-mesh of 1256 points in the irreducible part of the Brillouin zone
(30$\times$30$\times$30 in the full zone) was used to ensure accurate
density of states (DOS) information.

The electron localizability indicator was evaluated according to
Ref.~\onlinecite{Kohout04} with an ELI/ELF module implemented within
the FPLO program package.\cite{Ormeci06} The topology of ELI was
analyzed using the program Basin \cite{Kohout08} with consecutive
integration of the electron density in basins, which are bound by
zero-flux surfaces in the ELI gradient field. This procedure, similar
to the one proposed by Bader for the electron density \cite{Bader99}
allows to assign an electron count for each basin, providing
fingerprints of direct (covalent) interactions.

\begin{figure}[t,b]
     \includegraphics[width=0.45\textwidth]{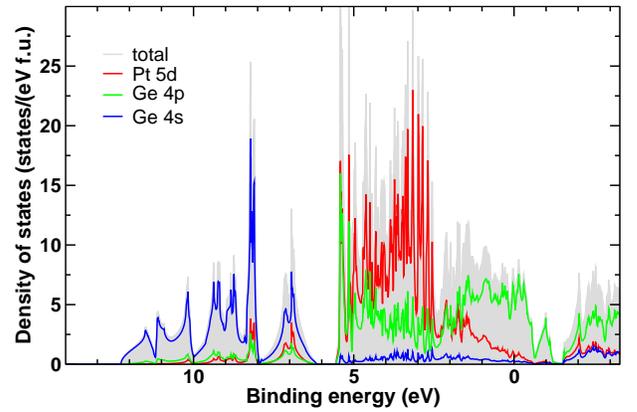}
     \vspace{-2mm} \caption{ Calculated total and atom resolved
     partial electronic density of states of SrPt$_4$Ge$_{12}$. The
     Fermi level is at zero energy.}  \label{theo_dos}\vspace{-3mm}
\end{figure}

Figure~\ref{theo_dos} shows the calculated DOS for SrPt$_4$Ge$_{12}$
($E_F$ is at zero binding energy). The valence band is almost
exclusively formed by Pt and Ge states. The low and featureless Sr DOS
indicates that it plays basically the role of a charge
reservoir. Further inspection of the DOS shows that the high-lying
states between about 6 and 12 eV binding energies originate
predominantly from Ge 4$s$ electrons, whereas the lower lying part of
the valence band is formed by Pt 5$d$ and Ge 4$p$ states. The Pt 5$d$
states essentially form a narrow band complex of approximately 3 eV
band width centered at about 4 eV binding energy. Our calculated DOS
is in good agreement with the previous results of Bauer et
al.\cite{Bauer07a}

\begin{figure}[t,b]
     \includegraphics[width=0.45\textwidth]{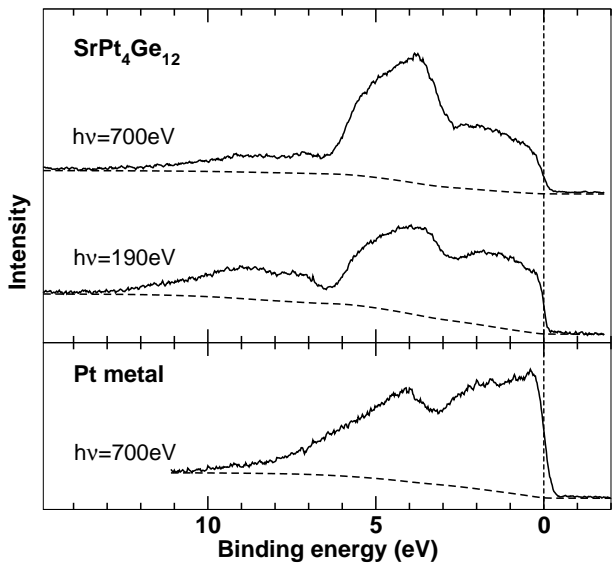}
     \vspace{-2mm} \caption{Valence band photoemission spectra of
       SrPt$_4$Ge$_{12}$ taken with a photon energy of
       h$\nu$\,=\,700\,eV (upper panel) and 190 eV (middle panel).  As
       reference, the valence band spectrum of elemental Pt metal
       taken at h$\nu$\,=\,700\,eV is also given (bottom panel).  The
       spectra are normalized with respect to their integrated
       intensities after subtracting an integral background indicated
       by the dashed curves.}
     \label{exp_xps}\vspace{-3mm}
\end{figure}

Figure~\ref{exp_xps} displays the valence band photoemission spectra
of SrPt$_4$Ge$_{12}$, taken with a photon energy of h$\nu$\,=\,700 eV
(upper panel) and 190 eV (middle panel), together with the reference
spectrum of elemental Pt metal (bottom panel). The spectra are
normalized to their integrated intensities, after an integral
background has been subtracted to account for inelastic scattering. All spectra
show a clear cut-off at $E_F$ (zero binding energy) consistent with
the systems being good metals. It is of no surprise that the 700 eV
spectrum of SrPt$_4$Ge$_{12}$ is very much different from that of Pt
since they are different materials. More interesting is that there are
also differences between the 700 eV and 190 eV spectra of
SrPt$_4$Ge$_{12}$ itself. This is caused by differences in the photon
energy dependence of the photo-ionization cross-section of the
relevant subshells forming the valence band, in this case, the Ge
$4s$, $4p$, and Pt $5d$.\cite{Yeh85} In fact, we chose those 190 eV
and 700 eV photon energies in order to make optimal use of the
cross-section effects for identifying the individual contributions of
the Ge and Pt states to the valence band as we will show in the next
sections. In particular, at 700 eV the Pt $5d$ cross-section is
calculated to be a factor 3.9 larger than that of the Ge 4p, while at
190 eV (close to the Cooper minimum for the Pt $5d$) it is equal or
even slightly smaller, i.e. a factor 0.92, see
Table~\ref{tabcross}. In other words, the 700 eV spectra is dominated
by the Pt $5d$ contribution while at 190 eV the contributions become
comparable.

\begin{table}
\caption{Calculated photo-ionization cross-sections per electron
($\sigma$ in Mb/e) for the Pt $5d$, Ge $4p$ and Ge $4s$ subshells,
from Yeh and Lindau.\cite{Yeh85}}\label{tabcross}
\vspace*{.2cm}
\begin{tabular}{c|c|c|c|c|c}
\hline\hline
$h\nu$ (eV)&$\sigma^{Pt5d}$&$\sigma^{Ge4p}$&$\sigma^{Ge4s}$&$\sigma^{Pt5d}$/$\sigma^{Ge4p}$&$\sigma^{Pt5d}$/$\sigma^{Ge4s}$\\\hline
190&0.0099&0.0108&0.021&0.92&0.47\\
700&0.0074&0.0019&0.0030&3.9&2.5\\
\hline\hline
\end{tabular}
\end{table}

The intensity $I$ of a normalized spectrum as depicted in
Figure~\ref{exp_xps} is built up from the Pt $5d$, Ge $4p$ and Ge $4s$
partial DOS ($\rho$), weighted with their respective photo-ionization
cross-sections ($\sigma$). This is formulated in equations 1 and 2
which take into account that the cross-sections at 190 eV photon
energy are different from those at 700 eV, respectively. The
proportionality factors $c_{190}$ and $c_{700}$, respectively, also
enter here since the absolute values for the photon flux and the
transmission efficiency of the electron energy analyzer are not
known. In addition, the constants $\alpha$, $\beta$, and $\gamma$ are
introduced to express the non-uniqueness in the calculation of the
weight of the Pt $5d$, Ge $4p$, and Ge $4s$ DOS, respectively, since
these depend (somewhat) on which calculational method has been used.

\begin{eqnarray}
I_{190} = c_{190}[\sigma_{190}^{Pt5d}\alpha\rho^{Pt5d} +
  \sigma_{190}^{Ge4p}\beta\rho^{Ge4p} +
  \sigma_{190}^{Ge4s}\gamma\rho^{Ge4s}]\\[.2cm]
I_{700} = c_{700}[\sigma_{700}^{Pt5d}\alpha\rho^{Pt5d} +
  \sigma_{700}^{Ge4p}\beta\rho^{Ge4p} +
  \sigma_{700}^{Ge4s}\gamma\rho^{Ge4s}]
\end{eqnarray}

Using the predictions of the band structure calculations as a guide,
we notice that the Ge $4s$ states hardly give a contribution in the
energy range between the Fermi level and 6 eV binding energy. In this
range, i.e. most relevant for the properties, the valence band seems
to be determined mostly by the Pt $5d$ and Ge $4p$ states. We now
analyze the experimental spectra along these lines. Using equations
(1) and (2), we can experimentally extract the Pt $5d$ and Ge $4p$ DOS
as follows:

\begin{eqnarray}
\rho^{Pt5d} \sim I_{700} - A \times I_{190}\\[.2cm]
\rho^{Ge4p} \sim I_{190} - B \times I_{700}
\end{eqnarray}
with $A$ =
($c_{700}/c_{190}$)$\times$($\sigma_{700}^{Ge4p}/\sigma_{190}^{Ge4p}$)
and $B$ =
($c_{190}/c_{700}$)$\times$($\sigma_{190}^{Pt5d}/\sigma_{700}^{Pt5d}$).
While the cross-sections $\sigma$ can be readily obtained from
Table~\ref{tabcross}, it would be an enormous task to determine
(experimentally) the ratio between $c_{700}$ and $c_{190}$. Thus,
since it is difficult to obtain directly an estimate for $A$ and $B$,
we use the product $AB$ given by
($\sigma_{190}^{Pt5d}/\sigma_{190}^{Ge4p}$)/($\sigma_{700}^{Pt5d}/\sigma_{700}^{Ge4p}$),
which can be calculated from Table~\ref{tabcross} to be about 0.92/3.9
= 0.24. Therefore, we vary the values for A and B under the constraint
that $AB$\,=\,0.24, searching for difference spectra which reproduce
both the Pt $5d$ and the Ge $4p$ DOS as obtained by the band structure
calculations.

\begin{figure}[t,b]
     \includegraphics[width=0.45\textwidth]{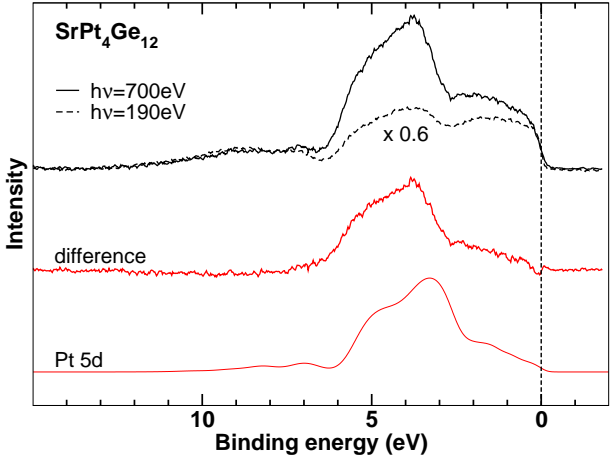}
     \vspace{-2mm} \caption{(Color online) Normalized and
       background-corrected valence band photoemission spectra of
       SrPt$_4$Ge$_{12}$ taken with a photon energy of
       $h\nu$\,=\,700\,eV (black solid line) and 190 eV (black dashed
       line). The 190 eV spectrum has been rescaled with a factor 0.6
       (see text). The difference spectrum (red solid line) is
       compared to the calculated Pt $5d$ DOS (red thin
       line).}  \label{diff700}\vspace{-3mm}
\end{figure}

\begin{figure}[t,b]
     \includegraphics[width=0.45\textwidth]{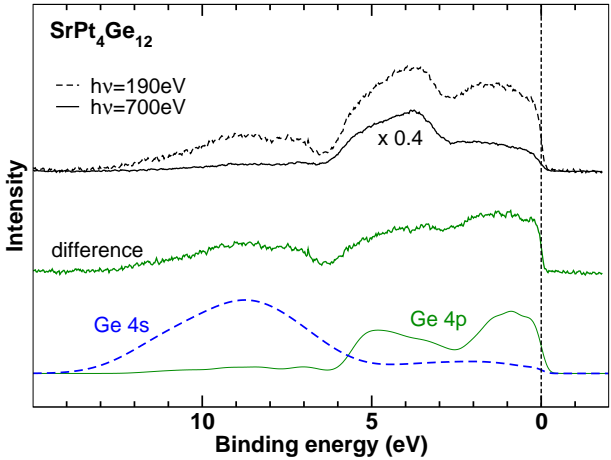}
     \vspace{-2mm} \caption{(Color online) Normalized and
       background-corrected valence band photoemission spectra of
       SrPt$_4$Ge$_{12}$ taken with a photon energy of h$\nu$=190 eV
       (black dashed line) and 700 eV (black solid line). The 700 eV
       spectrum has been rescaled with a factor 0.4. The difference
       spectrum (green solid line) is compared to the calculated Ge
       $4p$ DOS (green thin line) and Ge $4s$ DOS (blue dashed line).}
     \label{diff190}\vspace{-3mm}
\end{figure}

We find good results for $A$\,=\,0.6 and $B$\,=\,0.4 as displayed in
Figures~\ref{diff700} and \ref{diff190}. Focusing first at
Figure~\ref{diff700} in which the rescaled 190 eV spectrum is
subtracted from the 700 eV one, we can observe that the difference
spectrum resembles very much that of the calculated Pt $5d$ DOS. Here
the latter has been broadened to account for the experimental
resolution and lifetime effects. Interestingly, the main Pt intensity
is positioned at around 3-6 eV binding energies and its weight near
the $E_F$ region is very small. The experiment fully confirms this
particular aspect of the theoretical prediction, which is important
for the modeling of the superconducting properties as discussed
above. We would like to note that the main peak of the calculated Pt
$5d$ DOS is positioned at a somewhat lower binding energy than that of
the experimental difference spectrum. Similar small deviations have
been observed in other intermetallic materials\cite{Gegner06,Gegner08}
and can be attributed to the inherent limitations of mean-field
methods like the LDA to calculate the dynamic response of a system.

Figure~\ref{diff190} shows the difference of the spectrum taken at 190
eV and the rescaled spectrum at 700 eV. This experimental difference
spectrum reveals structures which can be divided into two major energy
regions: the first region extends from 0 to 6 eV binding energy, and
the second from 6 to 12 eV. For the first region we can make a
comparison with the calculated Ge $4p$ DOS, since the calculated Ge
$4s$ contribution is negligible as explained above. We obtain a very
satisfying agreement between the experiment and theory for the Ge
$4p$ states. In particular, we would like to point out that the experiment
confirms the strong presence the Ge $4p$ states in the vicinity of the
Fermi level $E_F$. Looking now at the second region, we see that the
calculated Ge $4s$ also reproduces nicely the experimental difference
spectrum. Here we remark that the calculated Ge $4p$ has a negligible
contribution in this region.

\begin{figure}[t,b]
     \includegraphics[width=0.45\textwidth]{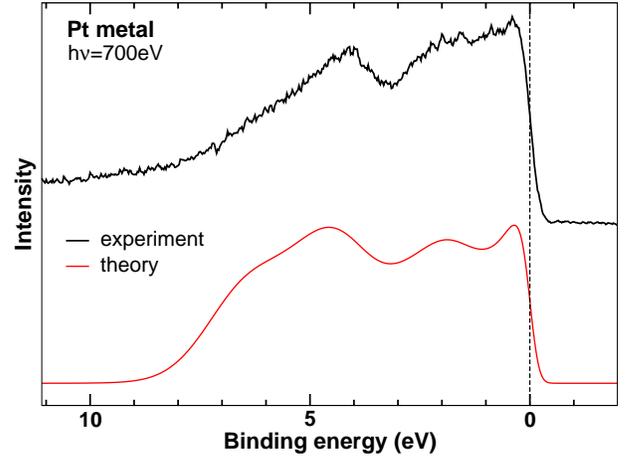}
     \vspace{-2mm} \caption{(Color online) Valence band photoemission
     spectrum of elemental Pt metal taken with a photon energy of
     $h\nu$\,=\,700\,eV (black-solid line) and the calculated Pt $5d$ DOS
     (red-solid line).}
     \label{PtMetal}\vspace{-3mm}
\end{figure}

As a further check we also perform a comparison between the
experimental photoemission spectrum of elemental Pt metal and the
corresponding calculated Pt $5d$ DOS. The result is shown in
Figure~\ref{PtMetal}. Also here we find satisfying agreement
between experiment and theory. The Pt $5d$ states range from 9 eV
binding energy all the way up to $E_F$. Clearly, the high Fermi
cut-off in Pt metal is formed by these Pt $5d$ states, in strong
contrast to the SrPt$_{4}$Ge$_{12}$ case.

\begin{figure}[t,b]
     \includegraphics[width=0.40\textwidth]{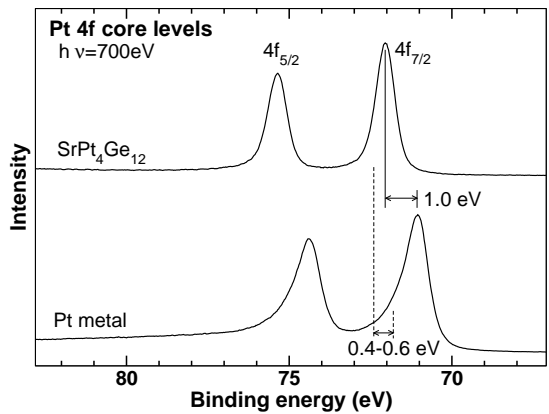}
     \vspace{-2mm} \caption{Pt 4$f$ core level photoemission spectra
     of SrPt$_4$Ge$_{12}$ (top) and elemental Pt metal (bottom). Solid
     vertical lines represent the peak positions of the $4f_{7/2}$
     levels, dashed vertical lines the center of gravity positions
     (see text).}  \label{CoreLevels4f}\vspace{-3mm}
\end{figure}

Figure \ref{CoreLevels4f} shows the Pt 4$f$ core levels of
SrPt$_4$Ge$_{12}$ (top) and elemental Pt metal (bottom). The spectra
exhibit the characteristic spin-orbit splitting giving the 4$f_{5/2}$
and 4$f_{7/2}$ peaks. For SrPt$_4$Ge$_{12}$, the peak positions for
the 4$f_{5/2}$ and 4$f_{7/2}$ are 75.4 and 72.1 eV binding energy,
respectively. For Pt metal, the values are 74.4 and 71.1 eV,
respectively. The spin orbit splitting is thus 3.3 eV for both
materials. This compares well with the calculated spin-orbit splitting
of about 3.45 eV for both compounds from the LDA calculations.

Remarkable is that the SrPt$_4$Ge$_{12}$ 4$f$ peaks are shifted by 1
eV to higher binding energies in comparison to those of Pt
metal. Similar shifts have also been observed in other noble metal
intermetallic compounds,\cite{Franco03,Gegner06,Gegner08} indicating a
more dilute electron density around the noble metal sites. To compare
this chemical shift to LDA calculations, one has to take into account
that LDA does not incorporate many body effects of the final state,
manifesting in the asymmetric line shape of the spectra as we will
discuss below in more detail. But it can be shown \cite{Lundqvist68}
that final state effects do not alter the average energy of the
spectrum. If we determine the center of gravity of the 4$f_{7/2}$, we
find a binding energy of 72.4 eV for SrPt$_4$Ge$_{12}$ and
71.9$\pm$0.2\,eV (indicated by dashed lines in Figure
\ref{CoreLevels4f}) for Pt metal, resulting in a chemical shift of
0.4\,eV to 0.6\,eV. This agrees well with the shift obtained from our
band structure calculations which amounts to 0.43 eV.

One can clearly observe that the line shape of the core levels in
SrPt$_4$Ge$_{12}$ is narrower and not as asymmetric as in the case of
Pt metal. An asymmetry in the line shape is caused by the presence of
electron-hole pair excitations upon the creation of the core hole,
i.e. screening of the core hole by conduction band electrons, and can
be well understood in terms of the Doniac-Sunjic
theory.\cite{Doniach70} The strong asymmetry of the 4$f$ of Pt metal
can therefore be taken as an indication for the high DOS with Pt
character at $E_F$.\cite{Huefner75} The rather symmetric line shape
of the 4$f$ of SrPt$_4$Ge$_{12}$, on the other hand, indicates very low
DOS at $E_F$. All this confirms the results of the valence band
measurements: the main intensity of the Pt 5$d$ band is at 3-6 eV
binding energies, strongly reducing its weight at $E_F$.

In conclusion, we find excellent agreement between the measured
photoemission spectra and the LDA band structure calculations for
SrPt$_4$Ge$_{12}$. This confirms the picture of the chemical bonding
analysis yielding rather deep lying Pt 5$d$ states which only partially form
covalent bands with the Ge 4$p$ electrons. In turn, the states at the
Fermi level that are relevant for the superconducting behavior of this
compound can be firmly assigned to Ge 4$p$ electrons. This study
provides strong support that band theory is a good starting point for
the understanding of the electronic structure of the
$M$Pt$_4$Ge$_{12}$ ($M$ = Sr, Ba, La, Pr, Th) material class, and is
thus of valuable help in the search for new compositions with higher
superconducting transition temperatures.


\end{document}